\documentclass[letterpaper, 10 pt, conference, twocolumn]{ieeeconf}

\IEEEoverridecommandlockouts                              

\usepackage{graphics} 
\usepackage{epsfig} 

\usepackage{times} 
\usepackage{amsmath} 
\usepackage{amssymb}  
\usepackage{dsfont}
\usepackage{booktabs}
\usepackage{mathtools}

\newtheorem{definition}{Definition}

\usepackage{nicefrac}       
\usepackage{microtype}      
\usepackage{tabto}
\usepackage{xcolor}

\usepackage{enumerate}
\usepackage{caption}
\usepackage{subcaption}
\usepackage{stfloats}

\usepackage[notocbasic]{nomencl}
\makenomenclature
\usepackage{etoolbox}

\usepackage{algorithm}
\usepackage[noend]{algorithmic}

\usepackage[colorlinks=true,linkcolor=blue]{hyperref}       
\usepackage{url}            

\usepackage{xpatch}
\xpatchcmd{\thenomenclature}{%
  \section*{\nomname}
}{
}{\typeout{Success}}{\typeout{Failure}}

\newcommand{\EE}{\mathbb{E}}

\newcommand{\RR}{\mathbb{R}}

\newcommand{\bbS}{\mathbb{S}}

\newcommand{\cE}{\mathcal{E}}
\newcommand{\cF}{\mathcal{F}}
\newcommand{\cG}{\mathcal{G}}
\newcommand{\cH}{\mathcal{H}}

\newcommand{\cK}{\mathcal{K}}
\newcommand{\cL}{\mathcal{L}}

\newcommand{\cR}{\mathcal{R}}

\newcommand{\tp}{\intercal}

\newcommand{\Let}{\coloneqq}

\DeclareMathOperator{\vect}{vec}

\DeclareMathOperator{\Tr}{Tr}

\newcounter{mytempeqncnt}  

\title{\LARGE \bf
Policy Iteration for Multiplicative Noise Output Feedback Control
}

\author{Benjamin Gravell \quad Matilde Gargiani\quad John Lygeros \quad Tyler H. Summers
\thanks{B. Gravell and T. Summers are with the Control, Optimization, and Networks Lab, University of Texas at Dallas (email: \{benjamin.gravell\}, \{tyler.summers\}@utdallas.edu). M. Gargiani and J. Lygeros are with the Automatic Control Laboratory, ETH Z\"urich, Switzerland. This material is based on work supported by the United States Air Force Office of Scientific Research under award number FA2386-19-1-4073, the National Science Foundation under award number ECCS-2047040, and the European Research Council (ERC) under the OCAL project 787845.}
}

\begin{document}

\maketitle
\thispagestyle{empty}
\pagestyle{empty}

\begin{abstract}
We propose a policy iteration algorithm for solving the multiplicative noise linear quadratic output feedback design problem. The algorithm solves a set of coupled Riccati equations for estimation and control arising from a partially observable Markov decision process (POMDP) under a class of linear dynamic control policies. 
We show in numerical experiments far faster convergence than a value iteration algorithm, formerly the only known algorithm for solving this class of problem. 
The results suggest promising future research directions for policy optimization algorithms in more general POMDPs, including the potential to develop novel approximate data-driven approaches when model parameters are not available.
\end{abstract}

\section{Introduction}
Multiplicative noise models can be used to represent myriad phenomena where noise or uncertainty depends on the system state, input, or output.
These models have a long history in control theory \cite{wonham1967optimal} 
and have been utilized in the context of networked control systems \cite{sinopoli2004kalman} 
robots with distance-dependent sensors such as lidar and optical cameras \cite{dutoit2011robot}, turbulent fluid flow \cite{lumley2007stochastic}, climate dynamics \cite{majda1999models}, biological sensorimotor systems \cite{todorov2005stochastic}, neuronal brain networks \cite{breakspear2017dynamic}, portfolio optimization and financial markets \cite{primbs2007portfolio} 
power grids with stochastic inertia \cite{guo2019a}, and aerospace systems \cite{gustafson1975design}.
Recently, these models have also been used to represent parametric uncertainty and promote robustness in data-driven control and machine learning via, e.g., bootstrapping \cite{gravell2020pmlr}, domain randomization \cite{tobin2017domain}, and dropout \cite{srivastava2014dropout}. 

Designing optimal output feedback controllers for multiplicative noise dynamical systems is particularly challenging because, in marked contrast to classical LQG/$\mathcal{H}_2$ and $\mathcal{H}_\infty$ control design, there is no separation between estimation and control. In particular, a canonical linear quadratic problem features a set of Riccati equations involving cost and covariance matrices for the state and state estimate that are coupled by the multiplicative noise. These equations cannot be solved by direct methods for decoupled Riccati equations \cite{bittanti2012riccati}. The only known solution method is an iterative algorithm akin to value iteration, described in \cite{dekoning1992}. 


Policy iteration has been studied extensively in general settings for dynamic programming and reinforcement learning and is closely related to the Newton method \cite{puterman1979convergence,bertsekas2021lessons}. Policy iteration and the Newton method have also been studied extensively for solving the single generalized Riccati equation arising in the linear quadratic state-feedback setting in \cite{kleinman1969,damm2004rational}. However, there is far less known about policy iteration and the Newton method for output feedback control or partially observable Markov decision processes (POMDPs). Various approximate policy iteration schemes have been studied to a limited extent in deterministic and additive noise output feedback linear quadratic problems in \cite{lewis2011,rizvi2018output,yaghmaie2022} and for tabular and nonlinear POMDPs in \cite{hansen1998pomdp, gao2016output}.
None of these address the coupled Riccati equations arising in the output-feedback multiplicative noise setting.

Our main contribution is to propose a policy iteration algorithm for multiplicative noise output feedback control design, which solves a set of coupled Riccati equations for estimation and control.
We show in numerical experiments far faster convergence than a value iteration algorithm, the only known algorithm for solving this class of problem. 
We provide an open source implementation of our algorithm to facilitate further research and reproducibility.
The results suggest promising future research directions for policy iteration and data-driven output feedback control algorithms for the multiplicative noise problem and other more general POMDPs.

The paper is organized as follows. \S\ref{sec:problem_formulation} formulates an optimal output feedback control problem for systems with multiplicative noise. \S\ref{sec:optimal_control} describes the coupled Riccati equations that determine an optimal policy in a form that motivates policy iteration. Our policy iteration algorithm is proposed in \S\ref{sec:policy_iteration}. \S\ref{sec:numerical_experiments} presents the numerical experiments, and \S\ref{sec:conclusion} concludes.

\textbf{Notation:}
Denote the set of real-valued $n \times m$ matrices as $\RR^{n \times m}$ and the set of $n \times n$ symmetric positive semidefinite matrices as $\bbS^{n}_{+}$.
Denote the $n \times n$ all-zeros and identity matrices as $0_n$ and $I_n$, respectively.
For real-valued matrices $A$ and $B$, denote the transpose as $A^\tp$, the trace as $\Tr(A)$, and the Frobenius inner product as $\langle A, B \rangle \Let \Tr(A^\top B)$.

\section{Problem Formulation: Output Feedback Control with Multiplicative Noise} \label{sec:problem_formulation}
We consider output-feedback control of the discrete-time stochastic linear dynamical system
\begin{subequations}
\label{eq:true_system}
\begin{align}
    x_{t+1} &= A_t x_t + B_t u_t + w_t,  \vspace{-0.1\baselineskip} \label{eq:true_system1}\\
    y_t     &= C_t x_t + v_t,  \label{eq:true_system2}
\end{align}
\end{subequations}
where $x_t \in \RR^n$ is the system state, $u_t \in \RR^m$ is the control input, and $y_t \in \RR^p$ is the observed output.
The system matrices $(A_t, B_t, C_t)$ are time-varying random matrices decomposed as
\begin{align*}
    A_t &= A + \sum_{i=1}^a \alpha_{t, i} A_{\Delta, i}, \quad
    B_t = B + \sum_{i=1}^b \beta_{t, i} B_{\Delta, i}, \\
    C_t &= C + \sum_{i=1}^c \gamma_{t, i} C_{\Delta, i},
\end{align*}
where $A = \EE[A_t]$, $B = \EE[B_t]$, $C = \EE[C_t]$ are mean matrices, $\left\{ \alpha_{t, i} \right\}_{i=1}^a$, $\left\{ \beta_{t, i} \right\}_{i=1}^b$, $\left\{ \gamma_{t, i} \right\}_{i=1}^c$ are random scalars mutually independent and independent across time with zero mean and standard deviations $\left\{ \sigma_{A, i} \right\}_{i=1}^a$, $\left\{ \sigma_{B, i} \right\}_{i=1}^b$, $\left\{ \sigma_{C, i} \right\}_{i=1}^c$ respectively, and $\left\{ A_{\Delta, i} \right\}_{i=1}^a$, $\left\{ B_{\Delta, i} \right\}_{i=1}^b$, $\left\{ C_{\Delta, i} \right\}_{i=1}^c$ are pattern matrices constant across time specifying the directions in which the multiplicative noise acts.
This decomposition can be obtained by an eigendecomposition of the covariance structure of the $A_t$, $B_t$, $C_t$, treating the eigenvalues as $\left\{ \sigma_{A, i} \right\}_{i=1}^a$, $\left\{ \sigma_{B, i} \right\}_{i=1}^b$, $\left\{ \sigma_{C, i} \right\}_{i=1}^c$ and eigenvectors as $\left\{ A_{\Delta, i} \right\}_{i=1}^a$, $\left\{ B_{\Delta, i} \right\}_{i=1}^b$, $\left\{ C_{\Delta, i} \right\}_{i=1}^c$ as discussed e.g. in \cite{gravell2021tac}.
The process and observation noises $w_t$ and $v_t$, respectively, are independent across time, have mean zero, and have covariance
\begin{align*}
    \EE \left[ \begin{bmatrix} w_t \\ v_t \end{bmatrix} \begin{bmatrix} w_t \\ v_t \end{bmatrix}^\tp \right] = 
    \begin{bmatrix}
    W_{xx} & W_{xy} \\
    W_{yx} & W_{yy}
    \end{bmatrix}
    = W \succ 0.
\end{align*}
The initial state $x_0$ is a random vector drawn from a distribution with mean zero and covariance $X_0$.
We consider convex quadratic stage cost in the states and inputs
\begin{align*}
    \ell(x_t, u_t) = 
    \begin{bmatrix} x_t \\ u_t \end{bmatrix}^\tp
    \begin{bmatrix}
    Q_{xx} & Q_{xu} \\
    Q_{ux} & Q_{uu}
    \end{bmatrix}
    \begin{bmatrix} x_t \\ u_t \end{bmatrix}, \quad 
    Q \succ 0.
\end{align*}
This yields the infinite-horizon average-cost multiplicative-noise linear quadratic output feedback control problem
\begin{subequations}
\label{eq:mlqc_original}
\begin{align}
    \min_{\pi} J(\pi) \Let & \lim_{T \to \infty} \frac{1}{T} \EE \left[ \sum_{t=0}^{T-1} \ell(x_t, u_t) \right] \label{eq:performance} \\
    & \text{subject to } \eqref{eq:true_system},
\end{align}
\end{subequations}
where a control policy $u_t = \pi(y_{0:t}, u_{0:t-1})$ dependent on the input-output history $y_{0:t}:= [y_0,y_1,...,y_t], u_{0:t-1}:=[u_0,u_1,...,u_{t-1}]$ is to be designed, and expectation is taken with respect to all random quantities in the problem, namely $\left\{x_{0}, \{ A_t \}, \{ B_t \}, \{ C_t \}, \{ w_{t} \}, \{ v_{t} \}\right\}$. 

\subsection{Linear Dynamic Control Policies}
A \emph{linear dynamic controller} is a widely used class of policy which combines a linear state estimator with a linear state estimate feedback in the form
\begin{subequations}
\label{eq:linear_dynamic_compensator}
\begin{align}
    \hat{x}_{t+1} &= F \hat{x}_t + L y_t, \\
    u_t &= K \hat{x}_t. 
\end{align}
\end{subequations}
The initial state estimate is chosen as $\hat{x}_0 = 0$, since the initial state has mean zero.
Such a controller is fully specified by the triple $(F, K, L)$ where $K \in \RR^{m \times n}$ is the control gain, $L \in \RR^{n \times p}$ is the state estimator gain, and $F \in \RR^{n \times n}$ is the closed-loop model matrix. 
We will consider the problem \eqref{eq:mlqc_original} with optimization over this special class of linear dynamic controllers.
In the classical LQG setting with additive noise-only  (where $A_t = A$, $B_t = B$, $C_t = C$), the optimal policy is indeed a linear dynamic controller of the class \eqref{eq:linear_dynamic_compensator}.
However, in the multiplicative noise setting this class may not be optimal.\footnote{Due to the multiplicative noise, the state distribution is non-Gaussian even when all primitive distributions are Gaussian, so the Kalman filter is not necessarily the optimal state estimator. To our knowledge, it has not been demonstrated whether there exist conditions under which a nonlinear controller outperforms the optimal linear dynamic controller for the problem \eqref{eq:mlqc_original}, but this is beyond our scope here.} 
Nevertheless, the work of \cite{dekoning1992} shows that restricting attention to \eqref{eq:linear_dynamic_compensator} admits useful stability characterizations and optimal control synthesis equations.


\subsection{Closed-Loop Dynamics and Performance Criterion}

\begin{figure*}[!b]
\normalsize
\setcounter{mytempeqncnt}{\value{equation}}
\setcounter{equation}{12}  
\hrulefill
\vspace*{4pt}
\begin{subequations}
\label{eq:qfun_ops}
\begin{align}
    \cG(X)
    &\Let 
    \begin{bmatrix}
    Q_{xx} & Q_{xu} \\
    Q_{ux} & Q_{uu}
    \end{bmatrix}
    + 
    \begin{bmatrix}
    A^\tp P A & A^\tp P B \\
    B^\tp P A & B^\tp P B
    \end{bmatrix} 
    +
    \begin{bmatrix}
    \sum_{i=1}^{a} \sigma^2_{A, i} A_{\Delta, i}^\tp P A_{\Delta, i} & 0 \\
    0 & \sum_{i=1}^{b} \sigma^2_{B, i} B_{\Delta, i}^\tp P B_{\Delta, i}
    \end{bmatrix} \nonumber \\
    & \quad + 
    \begin{bmatrix}
    \sum_{i=1}^{a} \sigma^2_{A, i} A_{\Delta, i}^\tp \hat{P} A_{\Delta, i} + \sum_{i=1}^{c} \sigma^2_{C, i} C_{\Delta, i}^\tp \cL(X)^\tp \hat{P} \cL(X) C_{\Delta, i}   & 0 \\ 0 & \sum_{i=1}^{b} \sigma^2_{B, i} B_{\Delta, i}^\tp  \hat{P} B_{\Delta, i}
    \end{bmatrix} \\
    \cH(X) 
    &\Let 
    \begin{bmatrix}
    W_{xx} & W_{xy} \\
    W_{yx} & W_{yy}
    \end{bmatrix}
    +
    \begin{bmatrix}
    A S A^\tp & A S C^\tp \\
    C S A^\tp & C S C^\tp
    \end{bmatrix}     
    +
    \begin{bmatrix}
    \sum_{i=1}^{a} \sigma^2_{A, i} A_{\Delta, i} S A_{\Delta, i}^\tp & 0 \\
    0 & \sum_{i=1}^{c} \sigma^2_{C, i} C_{\Delta, i} S C_{\Delta, i}^\tp
    \end{bmatrix} \nonumber \\  
    & \quad +
    \begin{bmatrix}
    \sum_{i=1}^{a} \sigma^2_{A, i} A_{\Delta, i} \hat{S} A_{\Delta, i}^\tp + \sum_{i=1}^{b} \sigma^2_{B,i} B_{\Delta, i} \cK(X) \hat{S}  \cK(X)^\tp B_{\Delta, i}^\tp & 0 \\
    0 & \sum_{i=1}^{c} \sigma^2_{C,i} C_{\Delta, i} \hat{S} C_{\Delta, i}^\tp
    \end{bmatrix}
\end{align}
\end{subequations}
\setcounter{equation}{\value{mytempeqncnt}}
\end{figure*}

We now develop some notation and expressions for the closed-loop dynamics and performance criteria, which will be useful later for describing a policy iteration algorithm.
Using a controller $(F, K, L)$, the closed-loop system dynamics are
\begin{align}
    \begin{bmatrix}
    x_{t+1} \\
    \hat{x}_{t+1}
    \end{bmatrix}
    =
    \begin{bmatrix}
    A_t & B_t K \\
    L C_t & F
    \end{bmatrix}
    \begin{bmatrix}
    x_{t} \\
    \hat{x}_{t}
    \end{bmatrix}
    + 
    \begin{bmatrix}
    I_n & 0 \\
    0 & L
    \end{bmatrix}
    \begin{bmatrix}
    w_{t} \\
    v_{t}
    \end{bmatrix}    \label{eq:closed_loop_dynamics_aug}
\end{align}
Denote the following augmented closed-loop matrices
\begin{subequations}
\label{eq:closed_loop_ldc} 
\begin{align}
    \Phi_t^\prime &\Let 
    \begin{bmatrix}
    A_t & B_t K \\
    L C_t & F
    \end{bmatrix}, \\
    Q^\prime &\Let
    \begin{bmatrix}
    I_n & 0 \\ 0 & K
    \end{bmatrix}^\tp
    \begin{bmatrix}
    Q_{xx} & Q_{xu} \\
    Q_{ux} & Q_{uu}
    \end{bmatrix}
    \begin{bmatrix}
    I_n & 0 \\ 0 & K
    \end{bmatrix}, \\
    W^\prime &\Let
    \begin{bmatrix}
    I_n & 0 \\ 0 & L
    \end{bmatrix}
    \begin{bmatrix}
    W_{xx} & W_{xy} \\
    W_{yx} & W_{yy}
    \end{bmatrix}
    \begin{bmatrix}
    I_n & 0 \\ 0 & L
    \end{bmatrix}^\tp.
\end{align}
\end{subequations}
Under the closed-loop dynamics \eqref{eq:closed_loop_dynamics_aug}, define the value and covariance matrices at time $t$ recursively by
\begin{subequations}
\label{eq:second_moment_dynamics}
\begin{align}
    P^\prime_{t+1}
    &=
    \EE \left[ {\Phi^\prime_t}^\tp P^\prime_{t} {\Phi^\prime_t}
    \right]
    +
    Q^\prime, \label{eq:second_moment_dynamics1} \\
    S^\prime_{t+1}
    &=
    \EE \left[ {\Phi^\prime_t} S^\prime_{t} {\Phi^\prime_t}^\tp
    \right]
    +
    W^\prime. \label{eq:second_moment_dynamics2}
\end{align}
\end{subequations}
with the initial value matrix $P^\prime_0 = Q^\prime$ and the initial second moment matrix $S^\prime_0 = \begin{bmatrix} X_0 & 0 \\ 0 & 0 \end{bmatrix}$.
In particular, $S^\prime_{t}$ is the second moment of the augmented state
\begin{align*}
    S^\prime_{t} 
    &=
    \EE \left[ 
    \begin{bmatrix}
    x_{t} \\
    \hat{x}_{t}
    \end{bmatrix}
    \begin{bmatrix}
    x_{t} \\
    \hat{x}_{t}
    \end{bmatrix}^\tp
    \right],
\end{align*}
while $P^\prime_{t}$ is related to the costs $\ell(x_t, u_t)$ through the relation
\begin{align*}
    \EE \left[ \sum_{t=0}^{T-1} \ell(x_t, u_t) \right] 
    =
    \langle P^\prime_T , S_0^\prime \rangle + \sum_{t=0}^T \langle P^\prime_t , W^\prime \rangle .
\end{align*}

We re-state some definitions from \cite{dekoning1992} that characterize asymptotic behavior of \eqref{eq:closed_loop_dynamics_aug}. 

\begin{definition}[Mean-square stability]
System \eqref{eq:closed_loop_dynamics_aug} is \emph{mean-square stable (ms-stable)} if there exists finite $S^\prime_{\infty} \in \bbS^{2n}_{+}$ such that $\lim_{k \rightarrow \infty} S^\prime_{t} = S^\prime_{\infty}$.
\end{definition}


\begin{definition}[Mean-square compensatability]
System \eqref{eq:closed_loop_dynamics_aug} is \emph{mean-square compensatable} if there exists a controller \eqref{eq:linear_dynamic_compensator} which renders the augmented system \eqref{eq:closed_loop_dynamics_aug} ms-stable.
\end{definition}


\textbf{Assumption.} System \eqref{eq:closed_loop_dynamics_aug} is \emph{mean-square compensatable}.

\noindent
Define the linear operators $\Psi(\cdot), \Gamma(\cdot): \bbS^n \to \bbS^n$ by
\begin{align*}
    \Psi(M) &\Let \EE [ {\Phi^\prime_t}^\tp M {\Phi^\prime_t} ] \\
    \Gamma(N) &\Let \EE [ {\Phi^\prime_t} N {\Phi^\prime_t}^\tp ].
\end{align*}
These operators govern the second moment dynamics \eqref{eq:second_moment_dynamics} as
\begin{align*}
    P^\prime_{t+1}
    &=
    \Psi(P^\prime_{t}) + Q^\prime, \\
    S^\prime_{t+1}
    &=
    \Gamma( S^\prime_{t} )
    +
    W^\prime.
\end{align*}
Accordingly, the ms-stability of the closed-loop system is characterized by the spectrum of the linear operator $\Psi(\cdot)$, or equivalently of $\Gamma(\cdot)$, since they share the same spectra. Namely, if the spectral radius $\rho(\Psi(\cdot)) < 1$ then the closed-loop system is ms-stable. 
From a computational viewpoint, the spectral radius of $\Psi(\cdot)$ is equal to the spectral radius of the associated matrix $\Psi$ that satisfies $\Psi M = \EE [ \vect({\Phi^\prime_t}^\tp M {\Phi^\prime_t}) ]$ for any symmetric matrix $M$, which takes the explicit form 
\begin{align}
    \Psi &= 
    {\Phi^\prime}^\tp \otimes {\Phi^\prime}^\tp
    + 
    \sum_{i=1}^a \sigma^2_{A,i} {A^\prime_{\Delta, i}}^\tp \otimes {A^\prime_{\Delta, i}}^\tp \label{eq:second_moment_matrix_builder} \\
    &+
    \sum_{i=1}^b \sigma^2_{B,i} {B^\prime_{\Delta, i}}^\tp \otimes {B^\prime_{\Delta, i}}^\tp
    +
    \sum_{i=1}^c \sigma^2_{C,i} {C^\prime_{\Delta, i}}^\tp \otimes {C^\prime_{\Delta, i}}^\tp, \nonumber
\end{align}
where 
\begin{align*}    
    \Phi^\prime 
    \Let 
    \EE[\Phi^\prime_t] 
    = 
    \begin{bmatrix}
    A & B K \\
    L C & F
    \end{bmatrix}, \quad
    A^\prime_{\Delta, i} \Let 
    \begin{bmatrix}
    A_{\Delta, i} & 0_n \\
    0_n & 0_n
    \end{bmatrix}, \quad \\
    B^\prime_{\Delta, i} \Let 
    \begin{bmatrix}
    0_n & B_{\Delta, i} K \\
    0_n & 0_n
    \end{bmatrix}, \quad
    C^\prime_{\Delta, i} \Let 
    \begin{bmatrix}
    0_n & 0_n \\
    L C_{\Delta, i} & 0_n
    \end{bmatrix}. 
\end{align*}
Note that a dual matrix $\Gamma$ can be computed by dropping all transpose marks (``${}^\tp$'') in \eqref{eq:second_moment_matrix_builder}.
With such ms-stability, the steady-state value matrix $P^\prime$ and the steady-state second moment $S^\prime$ are found by solving the discrete-time generalized Lyapunov equations
\begin{subequations}
\label{eq:dlyap_aug}
\begin{align}
    P^\prime &= \Psi(P^\prime) + Q^\prime, \label{eq:dlyap_aug1} \\ 
    S^\prime &= \Gamma(S^\prime) + W^\prime. \label{eq:dlyap_aug2}
\end{align}
\end{subequations}
With a slight abuse of notation, the performance criterion \eqref{eq:performance} can be expressed and computed as
\begin{align}
    J(F, K, L) = \langle P^\prime, W^\prime \rangle = \langle S^\prime, Q^\prime \rangle. \label{eq:performance_linear}
\end{align}
which is finite only when \eqref{eq:closed_loop_dynamics_aug} is \emph{mean-square compensatable}. 
Solving the generalized Lyapunov equations \eqref{eq:dlyap_aug} to evaluate the performance \eqref{eq:performance_linear} of a given policy will form a basic component of our policy iteration algorithm.
Also in preparation for the policy iteration algorithm, given $P^\prime, S^\prime$, define
\begin{subequations}
\label{eq:suboptimal_value_matrix}
\begin{align}
        P &= \begin{bmatrix} I_n & I_n \end{bmatrix} P^\prime \begin{bmatrix} I_n & I_n \end{bmatrix}^\tp, \\
        \hat{P} &= \begin{bmatrix} 0_n & I_n \end{bmatrix} P^\prime \begin{bmatrix} 0_n & I_n \end{bmatrix}^\tp , \\
        S &= \begin{bmatrix} I_n & -I_n \end{bmatrix} S \begin{bmatrix} I_n & -I_n \end{bmatrix}, \\
        \hat{S} &= \begin{bmatrix} 0_n & I_n \end{bmatrix} S^\prime \begin{bmatrix} 0_n & I_n \end{bmatrix}^\tp.
\end{align}
\end{subequations}
In particular, the second moment of state estimation error and state estimate are, respectively,
\begin{align*}
    S = \lim_{t \to \infty} \EE[( x_t - \hat{x}_t )( x_t - \hat{x}_t )^\tp], \quad
    \hat{S} = \lim_{t \to \infty} \EE[\hat{x}_t \hat{x}_t^\tp],
\end{align*}
and $P$, $\hat{P}$ have analogous interpretations in terms of the cost.

\section{Optimal Linear Feedback Control Design} \label{sec:optimal_control}
The optimal linear dynamic controller for the multiplicative noise problem \eqref{eq:mlqc_original} can be exactly computed by solving a set of coupled Riccati equations for estimation and control \cite{dekoning1992}. 
However, in the multiplicative noise setting \emph{there is no separation between estimation and control}, so the optimal controller gains $(F, K, L)$ must be jointly computed. Here we derive the equations in a form that facilitates the development of the policy iteration algorithm developed in the following section.
Specifically, the optimal gains can be computed by solving the coupled nonlinear matrix Riccati equation 
\begin{align}
    \cR(X) = 0 \label{eq:gen_riccati_tiny}
\end{align}
in $X = (P, \hat{P}, S, \hat{S})$ 
for the Riccati operator\\
$\cR: \bbS^{n} \times \bbS^{n} \times \bbS^{n} \times \bbS^{n} \to \bbS^{n} \times \bbS^{n} \times \bbS^{n} \times \bbS^{n}$ as
\begin{align} 
    \cR(X) \Let
    \begin{pmatrix}
    -P + \cG_{xx}(X) - \cG_{xu}(X) \cG_{uu}^{-1}(X) \cG_{ux}(X)  \\
    -\hat{P} + \cE(X) + \cG_{xu}(X) \cG_{uu}^{-1}(X) \cG_{ux}(X)  \\
    -S + \cH_{xx}(X) - \cH_{xy}(X) \cH_{yy}^{-1}(X) \cH_{yx}(X)  \\
    -\hat{S} + \cF(X) + \cH_{xy}(X) \cH_{yy}^{-1}(X) \cH_{yx}(X) 
    \end{pmatrix} \label{eq:coupled_riccati}
\end{align}
where we define the operators $\cG(X)$ and $\cH(X)$ in \eqref{eq:qfun_ops},
the closed-loop operators
\setcounter{mytempeqncnt}{\value{equation}}
\setcounter{equation}{\value{mytempeqncnt}+1}
\begin{align*}
    \cE(X) &\Let (A - \cL(X) C)^\tp \hat{P} (A - \cL(X) C) \\
    \cF(X) &\Let (A + B \cK(X)) \hat{S} (A + B \cK(X))^\tp
\end{align*}
and the gain matrix operators
\begin{subequations} 
\label{eq:coupled_riccati1_gains}
\begin{align}
    \cK(X) & \Let -\cG_{uu}^{-1}(X) \cG_{ux}(X), \\
    \cL(X) & \Let \cH_{xy}(X) \cH_{yy}^{-1}(X).
\end{align}
\end{subequations}
Note that $\cG$ and $\cH$  have a block $2 \times 2$ structure whose blocks are referred to with the same subscripts as in $Q$ and $W$ as appropriate.
These expressions can be derived from the matrix minimum principle \cite{athans1967} as in \cite{dekoning1992}.
Notice that the gains $\cK(X)$ and $\cL(X)$ depend only on the $\cG_{ux}$, $\cG_{uu}$, $\cH_{xy}$, $\cH_{yy}$ blocks of $\cG$ and $\cH$. 
Also notice that, unlike the $\cG_{xx}$ and $\cH_{xx}$ blocks, $\cG_{ux}$, $\cG_{uu}$, $\cH_{xy}$, $\cH_{yy}$ are not specified in terms of $\cK(X)$ and $\cL(X)$.
Hence, $\cK(X)$ and $\cL(X)$ can be computed explicitly in terms of $X$, and consequently so can $\cG(X)$ and $\cH(X)$.

The optimal controller is
\begin{align*}
    \big( A + B \cK(X^*) - \cL(X^*) C, \cK(X^*), \cL(X^*) \big)
\end{align*}
where $X^*$ solves \eqref{eq:coupled_riccati} \cite{dekoning1992}.
With this controller, the cost and second moment matrices satisfy
\begin{align*}
    P^\prime = 
    \begin{bmatrix}
    P + \hat{P} & - \hat{P} \\
    - \hat{P} & \hat{P}
    \end{bmatrix}, 
    \quad
    S^\prime =
    \begin{bmatrix}
    S + \hat{S} & \hat{S} \\
    \hat{S} & \hat{S}
    \end{bmatrix},
\end{align*}
and thus the optimal controller achieves the optimal cost
\begin{align*}
    J^* 
    &=
    \left\langle
    Q_{xx}, S 
    \right\rangle + 
    \left\langle
    \begin{bmatrix} I \\ K \end{bmatrix}^\tp
    \begin{bmatrix}
    Q_{xx} & Q_{xu} \\
    Q_{ux} & Q_{uu}
    \end{bmatrix} 
    \begin{bmatrix} I \\ K \end{bmatrix},
    \hat{S} 
    \right\rangle \\
    &=
    \left\langle
    W_{xx}, P 
    \right\rangle + 
    \left\langle
    \begin{bmatrix} I \\ -L \end{bmatrix}
    \begin{bmatrix}
    W_{xx} & W_{xy} \\
    W_{yx} & W_{yy}
    \end{bmatrix}
    \begin{bmatrix} I \\ -L \end{bmatrix}^\tp,
    \hat{P} \right\rangle.
\end{align*}

When the multiplicative noise terms are zero, the coupled Riccati equations reduce to the familiar two decoupled standard algebraic Riccati equations for optimal linear quadratic control and state estimation, which can be solved via several well-known methods such as the dynamic programming techniques of policy iteration and value iteration \cite{bertsekas2012dynamic}, convex semidefinite programming \cite{boyd1994linear}, and specialized direct linear algebraic methods \cite{bittanti2012riccati}. In contrast, the only known algorithm for solving the coupled Riccati equations \eqref{eq:coupled_riccati} is a value iteration-type algorithm, described in \cite{dekoning1992}.
This turns the Riccati equation \eqref{eq:gen_riccati_tiny} into a recursive update according to
\begin{align}
    X^{k+1} = X^k + \cR(X^k). \label{eq:value_iteration}
\end{align}
Convergence of \eqref{eq:value_iteration} was proved in \cite{dekoning1992} using a homotopic continuation argument that continuously deforms a multiplicative noise-free version of the problem (for which convergence of the value iteration algorithm is well-known) to the original multiplicative noise-driven problem.


\section{Policy Iteration for Multiplicative Noise Output Feedback Control} \label{sec:policy_iteration}
In this section, we describe a novel policy iteration algorithm to solve the coupled Riccati equations \eqref{eq:coupled_riccati}. 
Policy iteration is a well-known method for computing optimal policies in the full state-feedback setting, originating with \cite{kleinman1968iterative, hewer1971iterative} for linear quadratic problems.
It consists of two steps: (1) policy evaluation, where the value of the current policy is computed from the dynamics and cost; and (2) policy improvement, where the current policy is improved based on the policy evaluation. However, policy iteration is far less developed in the context of output feedback control problems or POMDPs. 

The form of the coupled Riccati equations in \eqref{eq:coupled_riccati} suggests a policy iteration algorithm analogous to the full state-feedback setting.
The operators $\cG(X)$ and $\cH(X)$ play a role analogous to the state-action value function (also called the Q-function) in reinforcement learning. In particular, for a given cost-covariance matrix variable $X$, the gain operators \eqref{eq:coupled_riccati1_gains} can be viewed as an update to the control and estimator gains that improves the corresponding value.
Combining this with a policy evaluation step from solving the generalized Lyapunov equations \eqref{eq:dlyap_aug} leads to a policy iteration algorithm for the multiplicative noise output feedback control problem detailed in Algorithm \ref{algorithm:policy_iteration_exact}.
Note that the initial policy must be ms-stabilizing so that solutions to \eqref{eq:dlyap_aug} exist; this is a standard assumption in policy iteration \cite{kleinman1968iterative, hewer1971iterative} and policy optimization algorithms \cite{gravell2021tac}.

\begin{algorithm}
\caption{Policy iteration for optimal dynamic output feedback of linear systems with multiplicative noise}
\begin{algorithmic}[1]
\label{algorithm:policy_iteration_exact}
    \REQUIRE ms-stabilizing policy $(A + B K^{0} - L^{0} C, K^{0}, L^{0})$, convergence threshold $\epsilon > 0$.
    \STATE Initialize $X^0 \! = \! (0, 0, 0, 0)$, $X^1 \! = \! (\infty, \infty, \infty, \infty)$, $k \! = \! 0$.
    \WHILE{$\|X^{k+1} - X^k \| > \epsilon$}
        \STATE \textbf{Policy Evaluation:} 
        Compute the value $X^k =(P^k, \hat{P}^k, S^k, \hat{S}^k)$ of the current policy $(A + B K^{k} - L^{k} C, K^{k}, L^{k})$ by finding the solutions ${P^\prime}^k$, ${S^\prime}^k$ to the Lyapunov equations \eqref{eq:dlyap_aug} and using the relations \eqref{eq:suboptimal_value_matrix}.
        \STATE \textbf{Policy Improvement:} 
        Update the policy according to
        \begin{align*}
        K^{k+1} = \cK(X^k) , \quad
        L^{k+1} = \cL(X^k)
        \end{align*}\;
        \vspace{-\baselineskip}
        \STATE $k \leftarrow k+1$
    \ENDWHILE
    \ENSURE Nearly optimal policy $(A + B K^{k} - L^{k} C, K^{k}, L^{k})$
\end{algorithmic}
\end{algorithm}

\section{Numerical Experiments} \label{sec:numerical_experiments}
The algorithms were implemented in Python and executed on a desktop PC with a quad-core Intel i7 6700K 4.0GHz CPU and 16GB RAM; no GPU computing was utilized.
Code supporting \S\ref{sec:numerical_experiments} is available in the GitHub repository
\begin{center}
\url{https://github.com/TSummersLab/policy-iteration-mlqc}
\end{center}

\subsection{Pendulum system} \label{sec:numerical_experiments_pendulum}
As a first example, we examined a particular system representing a forward Euler discretization of the continuous-time dynamics of a pendulum with torque actuation and control-dependent noise. The first and second states represent the angular position and velocity, respectively. 
The system dimensions were $n = 2, m = 1, p = 1$, $b = 1$, and $a = c = 0$, indicating $A_t$ and $C_t$ were unaffected by multiplicative noise.
The problem data were
\begin{align*}
    A &= 
    \begin{bmatrix}
    1.0 & 0.1 \\
    1.0 & 0.95
    \end{bmatrix}, \quad
    B = 
    \begin{bmatrix}
    0 \\ 0.1
    \end{bmatrix}, \quad
    B_{\Delta, 1} = 
    \begin{bmatrix}
    0 \\ 1.0
    \end{bmatrix},  \\
    C &= 
    \begin{bmatrix}
    1 & 0
    \end{bmatrix}, \quad
    \sigma_{B, 1} = 1.
\end{align*}
The penalty and additive noise covariance matrices were chosen as $Q = I$ and $W = \text{diag}(0, 0.01, 0.001)$ respectively. 
The multiplicative noise value $\sigma_{B, 1}$ was varied by scaling by a noise level factor $\eta \in [0, 1]$. 
Note that the system was open-loop ms-stable; therefore we selected as our initial policy the open-loop policy $(A, 0, 0)$.
The algorithms were terminated once the convergence criterion $\| X^{k} - X^{k-1} \| \leq 10^{-12}$ was achieved.

\begin{figure}[!htbp]
    \begin{subfigure}[b]{0.48\textwidth}
      \includegraphics[width=\textwidth]{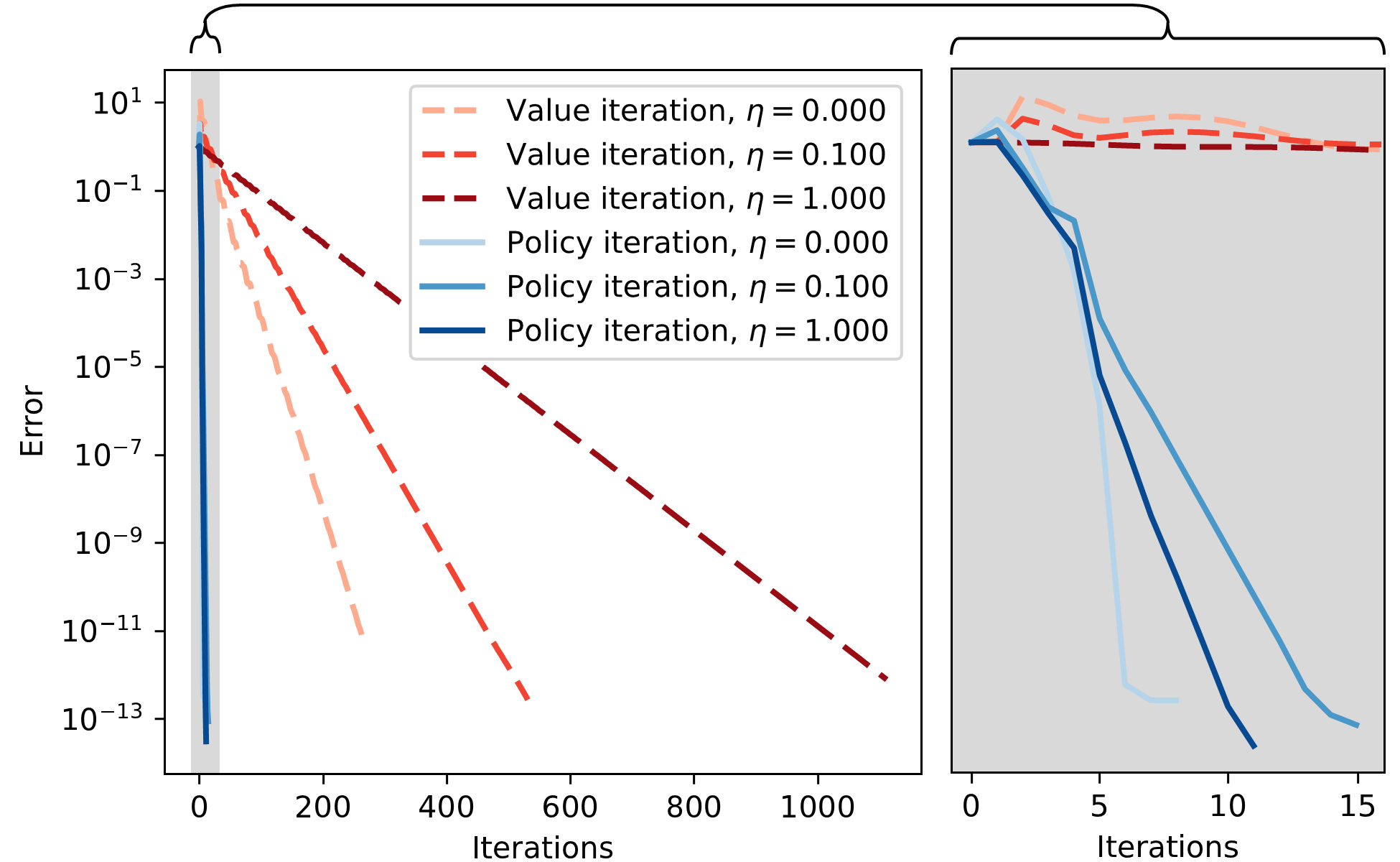}
      \caption{Relative error vs iteration count.}
    \end{subfigure} 
    \par\bigskip 
    \begin{subfigure}[b]{0.48\textwidth}
      \includegraphics[width=\textwidth]{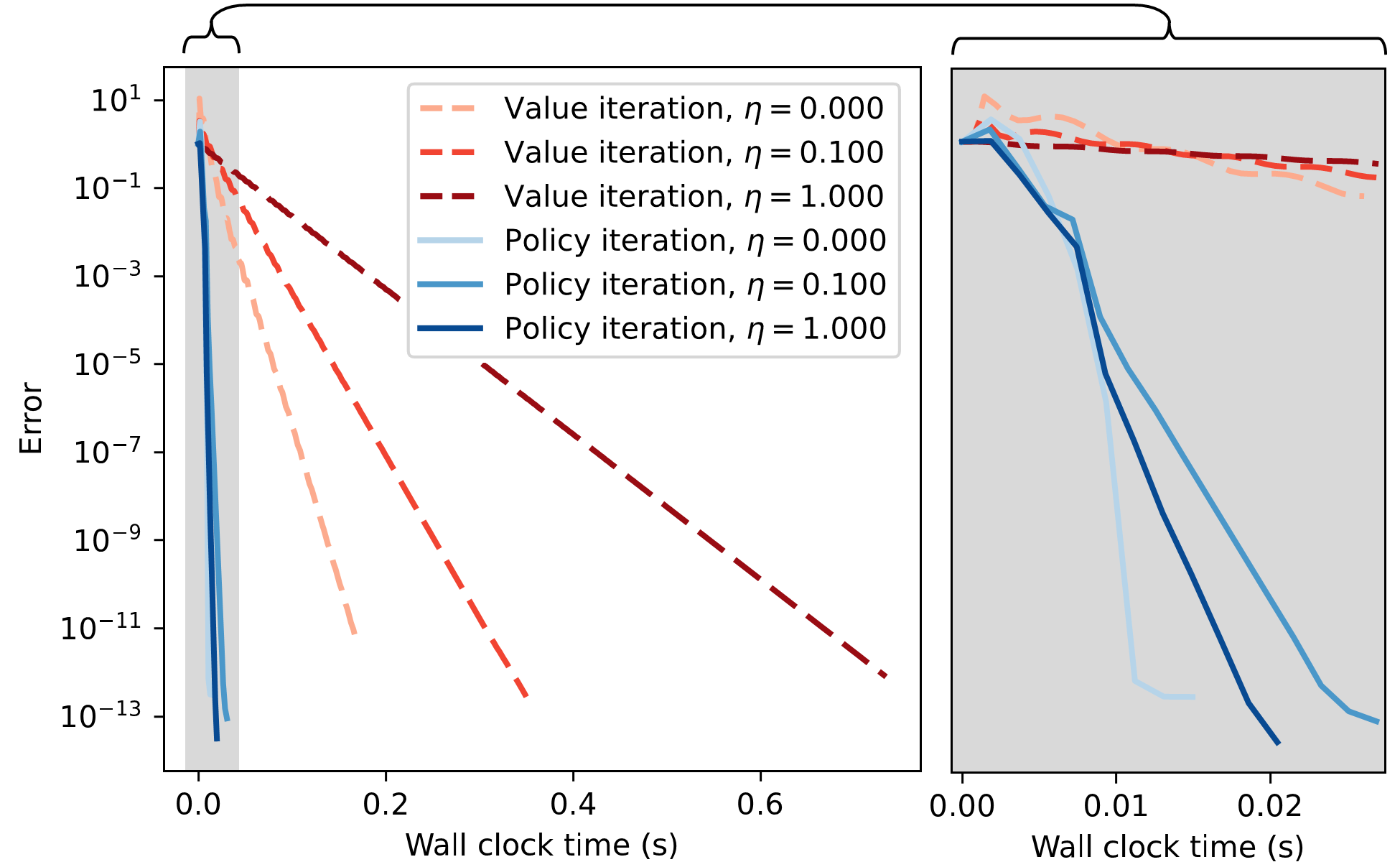}
      \caption{Relative error vs elapsed wall clock time.}
    \end{subfigure}
    \caption{Performance of value iteration and policy iteration for the pendulum system with various noise levels $\eta$.}
    \label{fig:pendulum_error}
  \end{figure}

Figure \ref{fig:pendulum_error} compares the policy iteration Algorithm \ref{algorithm:policy_iteration_exact} with the value iteration method of \cite{dekoning1992}.
The metric used for comparison was
\begin{align*}
    e^k &= \max \{ \delta(P^k), \delta(\hat{P}^k), \delta(S^k), \delta(\hat{S}^k)  \}, \\
    \text{where} \quad \delta(M^k) &= \| M^k - M^* \| / \| M^0 - M^* \|,
\end{align*}
which measures the error of the current solution relative to that of the initial solution.

The results demonstrate two important issues which arise and are addressed by the proposed policy iteration algorithm:
\begin{enumerate}
    \item The value iteration algorithm converges much more slowly compared to policy iteration, both in the multiplicative noise-free setting ($\eta = 0$) as well as in the multiplicative noise setting ($\eta > 0$).
    \item The presence of multiplicative noise causes value iteration to converge significantly more slowly, while the proposed policy iteration suffers a much less dramatic slowdown as the noise level $\eta$ increases.
\end{enumerate}
These observations apply both in terms of iteration count and wall clock time.

\subsection{Random systems} \label{sec:numerical_experiments_random}
Next, we report results for systems with randomly generated parameters.
The dimensions of each problem were chosen as $n = 2, m = 1, p = 1$ and $a = b = c = 1$.
The entries of $A$, $B$ $C$, $A_{\Delta, 1}$, $B_{\Delta, 1}$, $C_{\Delta, 1}$ were randomly drawn from a standard normal distribution. 
$A$ was scaled to have $\rho(A)$ drawn uniformly random between 0 and 1.
The variances $\sigma_{A, 1}^2$, $\sigma_{B, 1}^2$, $\sigma_{C, 1}^2$ were drawn uniform randomly between 0 and 1, then scaled such that the open-loop system with $(A, 0, 0)$ had $\sqrt{\rho(\Psi)} = 1$.
Finally, the variances $\sigma_{A, 1}^2$, $\sigma_{B, 1}^2$, $\sigma_{C, 1}^2$ were scaled by a factor $\eta$ drawn uniform randomly between 0 and 1.
Hence, the systems were always open-loop ms-stable by construction; therefore we selected as our initial policy the open-loop policy $(A, 0, 0)$.
The penalty and additive noise covariance matrices were chosen as $Q = I$ and $W = 0.01 I$ respectively.

\begin{figure}[!htbp]
    \centering
    \includegraphics[width=1\linewidth]{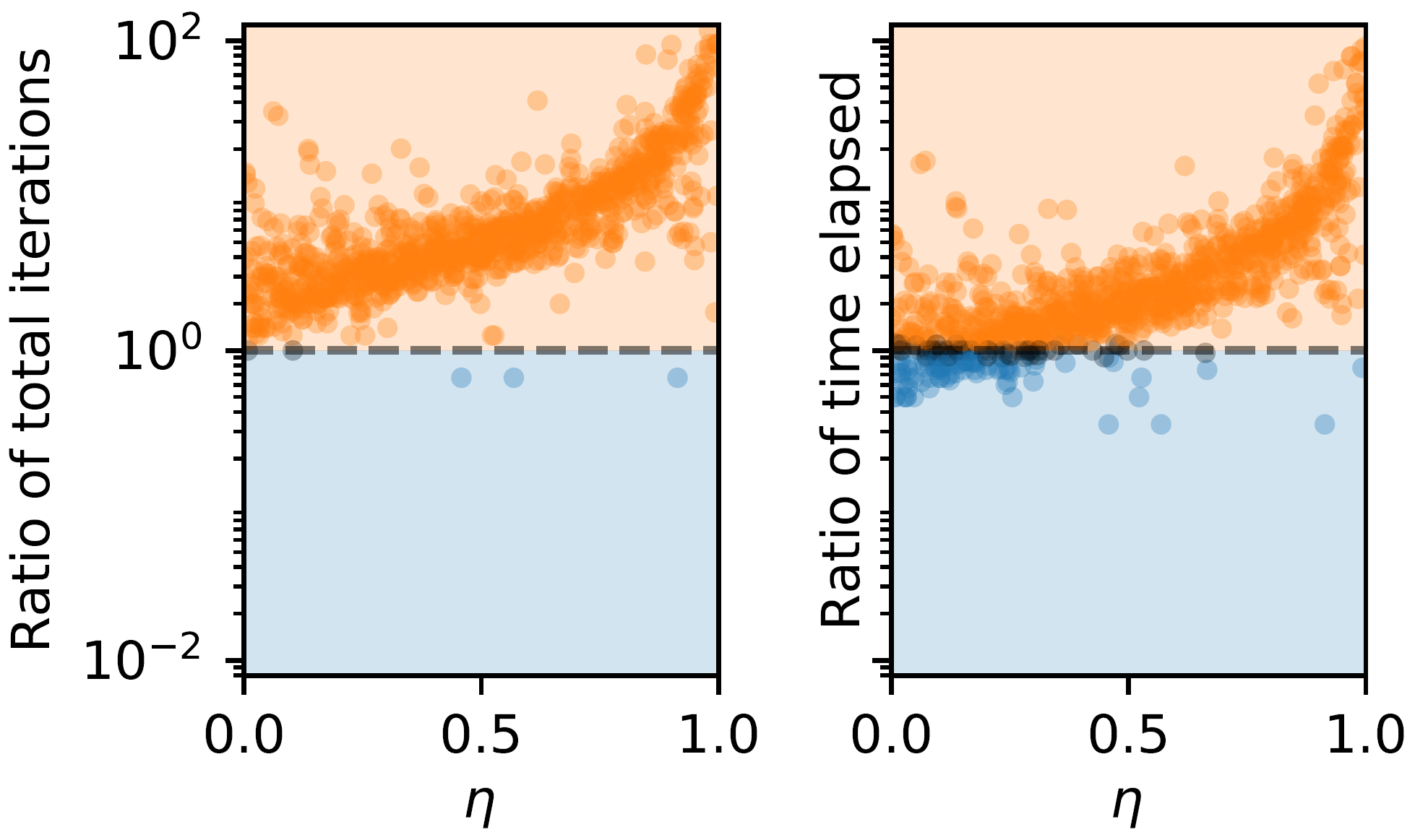}
    \caption{Ratio of total iterations and time elapsed using value iteration to that using policy iteration.}
    \label{fig:many_systems_vi_pi_both_scatter}
\end{figure}

The results are plotted in Figure \ref{fig:many_systems_vi_pi_both_scatter}, where each solved problem instance is represented by a scatter point.
It is evident that the proposed policy iteration algorithm converges in far fewer total iterations on almost every problem instance.
Due to greater per-iteration computation costs, the benefit in time elapsed is not as large for the policy iteration algorithm overall.
However, per-iteration costs can be significantly reduced by using specialized solvers for the generalized Lyapunov equations based on e.g. alternating-direction implicit preconditioning and Krylov subspaces \cite{damm2008direct} and other schemes that take advantage of sparsity in the linear matrix equation.
Nevertheless, the greatest benefit was achieved when the multiplicative noise was high relative to the maximum uncertainty that allows mean-square compensatability, which can be seen in the trend of Figure \ref{fig:many_systems_vi_pi_both_scatter}, mirroring the results observed for the pendulum system in \S\ref{sec:numerical_experiments_pendulum}.

The observations of \S\ref{sec:numerical_experiments_pendulum} and \S\ref{sec:numerical_experiments_random} together suggest that the least computationally costly method of solving \eqref{eq:mlqc_original} may be problem-instance dependent; a reliable indicator of when our policy iteration Algorithm \ref{algorithm:policy_iteration_exact} will outperform the value iteration \eqref{eq:value_iteration} remains to be discovered.

\section{Conclusions} \label{sec:conclusion}
We proposed a policy iteration algorithm for multiplicative noise optimal output feedback control design. Empirically it converges far faster than a value iteration algorithm, the only other known method for this problem.

Given the intimate connections between policy iteration and the Newton method, we believe there is a connection between our proposed algorithm and the Newton method (or some variation thereof) applied to the coupled Riccati equations \eqref{eq:coupled_riccati}. 
Interpreting our algorithm through a Newton lens would allow us to leverage rich theory surrounding the Newton method to obtain theoretical convergence results, in particular a quadratic rate of convergence as in \cite{kleinman1968iterative, hewer1971iterative, damm2001}. Furthermore, establishing a rate of convergence for the value iteration algorithm \eqref{eq:value_iteration}, which we conjecture is a linear rate, is, to our knowledge, also an open problem whose solution is required to establish theoretical relative performance claims. We leave this for future work.

Finally, our policy iteration algorithm opens the door to investigate novel approximate policy iteration algorithms that use input-output and state estimate data (instead of knowledge of system parameters) to approximately execute the policy evaluation and policy improvement steps. Many variations of approximate policy iteration have been explored in the full-state feedback setting; see e.g.
\cite{bradtke1994adaptive, krauth2019finite, gravell2021approximate}.
But there are many open problems, especially for approximate policy iteration in POMDPs. For example, it may be possible to use input-output and state estimate data generated from a given policy to form a least-squares estimate of the cost and covariance operators, and then perform an approximate policy improvement step based on this estimate. We will also explore this and other variations in future work.

\bibliographystyle{IEEEtran}
\bibliography{bibliography.bib}

\end{document}